\documentclass[prl,twocolumn,superscriptaddress,showpacs]{revtex4}

\usepackage[]{graphicx}

\newcommand{\ket}[1]{\left| #1 \right\rangle}

\begin{document}

\title{Topological phase for entangled two-qubit states}

\author{P\'erola Milman}
\email{Perola.Milman@lkb.ens.fr}
\affiliation{Laboratoire Kastler Brossel,
            D\'epartement de Physique de l'Ecole Normale Sup\'erieure, \\
            24 rue Lhomond, F-75231 Paris Cedex 05, France}
\affiliation{Coll\`ege de France, 11 place Marcelin-Berthelot, F-75005
            Paris, France}

\author{R\'emy Mosseri}
\email{mosseri@gps.jussieu.fr}
\affiliation{Groupe de Physique des Solides, Universit\'es Pierre et Marie Curie Paris 6 et Denis Diderot Paris 7 \\
2, Place Jussieu, 75251, Paris Cedex 05, France}

\date{\today}


\begin{abstract}
We present an unambiguous characterization of the rotation group
$SO(3)$ biconnectedness topology using  two-qubit maximally
entangled states. We show how to generate cyclic evolutions of these states, which are in one-to-one correspondence to closed paths in  $SO(3)$. The difference between the well known two classes of such paths translates into the
gain of a global phase of $\pi$ for one class and no phase change for the other. We  propose a simple quantum optics
interference experiment to demonstrate this topological phase shift.\end{abstract}

\maketitle

Two-qubit states are the simplest quantum mechanical systems
displaying entanglement. Maximally entangled states (MES) allow
for a clear and measurable distinction between classical and
quantum mechanical predictions, as exemplified by Bell type
inequalities \cite{Bell}. They also play a decisive role in
quantum information processes, like teleportation or dense coding
\cite{QuInfo}. In addition, it is known that any quantum computing
protocol involving many qubits can be implemented by concatenation
of one or two-qubit gates \cite{QuInfo}. This is why much
attention has been paid to two-qubit systems, their properties and
characterizations.

In the present work, we intend to show that two-qubit MES can also be used to display, even experimentally,
the well known double-connectedness of the $SO(3)$ rotation group. Since the latter is often evoked to
explain the minus sign multiplying a spin $1/2$ state after a $2\pi$ rotation, we shall first argue that
this is to some respect an ambiguous statement. Indeed, take one qubit
state $\ket{\Psi(t=0)}=\alpha \ket{0}+\beta \ket{1}$ subject to the
Hamiltonian $\hat H=\frac{\hbar \omega}{2} \hat \sigma_z$ (an effective magnetic field along the $z$ axis).
     At time $t$, the state
      reads $\ket{\Psi(t)}=e^{\frac{-i \omega t}{2}}\alpha \ket{0}+e^{\frac{i \omega t}{2}}\beta \ket{1}$. Let us turn to the Bloch sphere representation of $\ket{\Psi(t)}$.
        The three coordinates are given by the expectation values of the Pauli matrices. As is well known,
         the expectation value of the spin precesses around the effective magnetic field. At $t=2 \pi/\omega$ it
          has experienced a full $2\pi$ rotation, and a $\pi$ phase for the wave function. The $\pi$ phase has been
           clearly demonstrated on several systems, starting with beautiful experiment on spin $1/2$
            neutrons \cite{RauchWerner}. On general grounds, the change of global phase $\gamma$ under a cycle
             decomposes into its dynamical part, $\gamma_d$ (as derived from the time dependent Schr\"odinger equation)
              and its geometrical part, the Berry phase $\gamma_g$, whose origin relies on the Hilbert space peculiar
               geometry. The latter phase, $\gamma_g$, equals half of the area bounded by the representing part of
                the Bloch sphere \cite{Pancha}. The dynamical phase $\gamma_d$ is simply related to the
                 time average of the Hamiltonian. In the above case of one qubit precessing in a magnetic
                  field, we have $\gamma_d=-\pi \cos{\theta}$ and $\gamma_g=-\pi(1- \cos{\theta})$, so the expected
                  global phase $\gamma=\gamma_d+\gamma_g=-\pi$ for an initial state
                  with $|\alpha|=\cos{\theta/2}$ and $|\beta|=\sin{\theta/2}$. Let us remark here that
                  this $\pi$ phase is present even for $\theta=0$, where it is of pure dynamical nature (with no precession and therefore no rotation at all). We may
                  therefore question the geometrical interpretation of the $\pi$ phase. It would be more correct to state
                  that this phase has a geometrical and a dynamical component, being purely geometrical only
                  when $\theta=\pi/2$, when the initial state is orthogonal to the magnetic field. However, even in that
                  case, we find problems in relating the phase to the double connectedness of  $SO(3)$. Indeed, the latter
                  property relates paths on the $SO(3)$ manifold, stating that, under continuous deformations there are
                  only two different classes of paths, forming a two-element $Z_2$ first homotopy
                  group: $\Pi_1(SO(3))$. Note that two dimensional rotations ($SO(2)$), have a more complicated
                  structure: $\Pi_1(SO(2))=Z$ \cite{Nash}. The above experiment on a spin $1/2$ particle was done with a
constant magnetic field and therefore a constant rotation axis. A
natural question is the relation of the measured $\pi$ phase to
topological properties  of the rotation group: is the $\pi$ phase
related to the subtle nature of $SO(3)$ or to a property shared by
$SO(3)$ and its subgroup $SO(2)$, as it seems to be the case? In
the latter case, this phase gain is therefore not directly related
to the $SO(3)$ double connectedness, but rather to a multi
connectedness property shared by both groups.

In the following we aim to develop a non ambiguous relation
between the $SO(3)$ topology and the global phase of a wave
function. The idea is rather simple: we use the one-to-one map between
$SO(3)$ and the set of two qubit MES \cite{Mosseri:JPA2001}. A suitable set of
Hamiltonians that keep constant the degree of entanglement is constructed,
so that the two-qubit MES traces a path in $SO(3)$ when the
orientation of the Hamiltonians is changed. The evolution is such
that the dynamical phase vanishes while its geometrical
counterpart, in a sense that will be precised, also vanishes,
except at points where it abruptly changes to $\pi$. Note that such a $\pi$ phase, although not related to the
 $SO(3)$ topology, has already been discussed in the context of entangled states \cite{Geom2qubit}.

A pure general bipartite state can be written as
\begin{equation}\label{state}
\ket{\Psi}=\alpha\ket{00}+\beta\ket{01}+\gamma\ket{10}+\delta \ket{11},
\end{equation}
where $\alpha$, $\beta$, $\gamma$ and $\delta$ are complex
coefficients satisfying the normalization condition. The complex
concurrence ${\cal C}$ of this state is defined as ${\cal
C}=2(\alpha \delta- \beta \gamma)$ and its norm equals the standard concurrence $C$ for pure 2-qubits as
 defined by Wooters \cite{Wooters:PRL1998}. The concurrence $C$ is a
measure of the degree of entanglement for pure bipartite states,
assuming the value $C=1$ for maximally entangled states and
$C=0$ for product states. For the
case of MES, state (\ref{state}) can always be written (up to a global phase) in the
general form \cite{Mosseri:JPA2001}:
\begin{equation}\label{statemes}
\ket{\Psi}=\sqrt{\frac{1}{2}} \left( \alpha\ket{00}+\beta\ket{01}-{ \beta^*}\ket{10}+{  \alpha^*}\ket{11}\right ),
\end{equation}
where the star denotes complex conjugation. The Hilbert space of all the MES can thus be defined as $\Omega_{MES}=\left\{(\alpha,\beta) \in C^2 / |\alpha|^2+|\beta|^2=1, {\rm and } (\alpha,\beta) \sim (-\alpha,-\beta)\right \}={S^3 / Z_2}=SO(3)$, i.e., there is an one-to-one correspondance between $SO(3)$ and the MES. This represents a sphere (in $4$ d) whose opposite points are identified. By writing states in the form of Eq. (\ref{statemes}), we can see that each MES
can be denoted by a pair of complex numbers $(\alpha,\beta)$.

We will now study  closed trajectories in $SO(3)$,  corresponding to the time evolution of  MES subjected to
 precise Hamiltonians. The net result is that, depending on the homotopy classes to which the path belongs,
  the MES will acquire either a $0$ or $\pi$ phase. In the following we will show how this purely topological
   characteristics can be theoretically implemented and illustrated via an interference experiment.

\begin{figure}[h]
\center
\includegraphics[width=2.7in, height=2.5in]{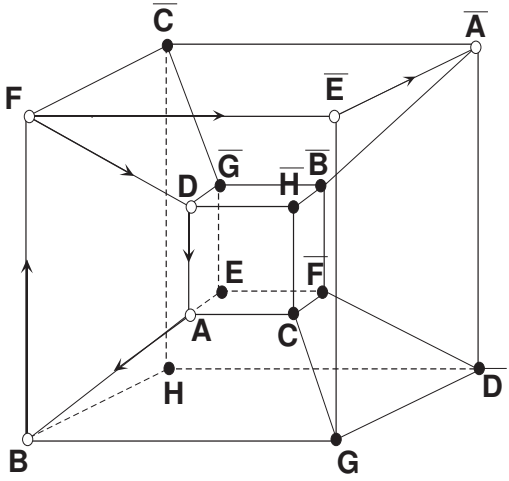}
\caption{Hypercube depicting the two topologically distinct
trajectories. The {\it plus} circuit, corresponding to no phase
change, is represented by the sequence of points $A\rightarrow B
\rightarrow F \rightarrow D \rightarrow A$. The {\it minus}
circuit, corresponding to the gain of a $\pi$ phase,  is the
represented by the sequence $A\rightarrow B \rightarrow F
\rightarrow \overline E \rightarrow \overline A$. } \label{fig1}
\end{figure}

Let's take as the initial MES the Bell state $(\alpha=1, \beta=0)$
\begin{equation}\label{init}
\ket{\Psi}=\sqrt{\frac{1}{2}}\left ( \ket{00}+\ket{11} \right ).
\end{equation}
In order to implement the cyclic evolution of state (\ref{init}),
we will consider simple polygonal trajectories in $SO(3)$: A sequence
of evolution operators  will act for a fixed finite time interval
on the states, until they reach back the initial state
(\ref{init}). A geometrical picture  illustrating this time
evolution can be given by the hypercube of $16$ vertices (belonging to $S^3$) depicted
in Fig. (\ref{fig1}). To get a discretized approximation of $SO(3)$ ($=S^3/Z_2$) we must identify opposite vertices, leaving altogether $8$ distinct points,  from $A$
to $H$ in Fig. (\ref{fig1}), and the edges connecting these points. They all  correspond to MES. The bar
 over the letters denote the
opposite points. Point $A$ is our
initial state (\ref{init})  and  point ${\overline A}$ is
identified to $-A$. The first class of trajectories that will be
studied starts at $A$ and finishes at the same point, making a
circuit in the polygon defined by points $A \rightarrow B
\rightarrow F \rightarrow  D \rightarrow  A$ in Fig. (\ref{fig1}).
This class will be called from now on the {\it plus} class, since
it is homotopic to the identity and does not imply in a global
phase change of the initial state. An example of the other class
of trajectory, the {\it minus} class, is given by the sequence of
points $A \rightarrow B \rightarrow F \rightarrow \overline E
\rightarrow \overline A$. It also starts at point $A$ but ends at
the point diametrically opposed to the initial one ($\overline
A$). Thus, there is a global phase of $\pi$ gained by the initial
state. Physically, this phase depends on the parity of the number of times the state crosses the space
  orthogonal to the initial one.

In order to keep the one-to-one relation between entangled states and $SO(3)$ elements, we will
consider, in our trajectories, evolution operators leaving the
complex concurrence ${\cal C}$ constant. Taking $\hbar \omega=1$, such unitary evolutions can be
realized by the Hamiltonian $\hat H= \frac{1}{2} {\vec n}. {\vec \sigma}$, whith ${\vec {\hat \sigma}}= \hat \sigma_x \vec x+ \hat \sigma_y \vec y+  \hat \sigma_z \vec z$, $\hat \sigma_x, \hat \sigma_y$
and $\hat \sigma_z$ and $\vec n=n_x \vec x+n_y \vec y+n_z \vec z$  is a normalized vector giving the orientation of the effective magnetic field. This Hamiltonian represents a rotation
acting on only one of the qubits. The evolution operator corresponding to this Hamiltonian has the form:
\begin{equation}\label{evop}
\hat U=e^{-i\hat H t}=
    \pmatrix{
            \cos{\frac{t}{2}}-in_z \sin{\frac{t}{2}}& -in_- \sin{\frac{t}{2}}\cr
          -in_+ \sin{\frac{t}{2}} & \cos{\frac{t}{2}}+in_z \sin{\frac{t}{2}}
            },
\end{equation}
where $n_{\pm}=n_x\pm in_y$.  Defining $\epsilon=e^{i\pi/4}/\sqrt{2}$,
 we have, for the points appearing in the hypercube of Fig. (\ref{fig1}):$A \rightarrow(1,0)$, $B \rightarrow( \epsilon,  \epsilon)$, $C\rightarrow( \epsilon^*,  \epsilon^*)$, $D\rightarrow (\epsilon, - \epsilon)$, $E \rightarrow (\epsilon^*, - \epsilon^*)$, $F\rightarrow(i, 0)$, $G \rightarrow(0, 1)$, $H\rightarrow(0, i)$. By changing the interaction time $t$ and the direction of rotation of our two-level system, i.e., vector $\vec n$, we are able to reach,
departing from point $A$, all points in $SO(3)$, representing all possible MES. Notice that all rotations are performed
on {\it one particle only}. In order to find the exact Hamiltonians that produce each class of trajectory, we
 will chose, for reasons that will become clear in the following, to perform the first two   rotations in the first
 qubit and the last two in the second qubit. Thus, the total operator
  acting on our states can be written in the form $\hat U \otimes  Id$ for the first half of the trajectory and
  $Id \otimes \hat U$ for the second one. One can easily check that in both cases the dynamical phases vanish.
  As for the geometrical phase, it can also be defined for open paths, as proposed by Pancharatnam \cite{Pancha}.
   This phase also vanishes in the present case, except when the state crosses the space orthogonal to the initial state,
    where it abruptly changes by $\pi$.

The parameters in equation (\ref{evop}) can be easily found by comparing this matrix to the necessary rotations to
 implement each
part of the circuits. By doing so, we find that each part of the trajectory takes a time $t=2\pi/3$ and the only difference
in the Hamiltonians is in the direction to which $\vec n$ points. These orientations $(n_x, n_y, n_z)$ read:
\vspace{0.5cm}
\begin{center}
\begin{tabular}{|l|l|}
\hline
$A \rightarrow B$ & $ \sqrt{1/3}(-1,-1,-1)$       \\ \hline
 $B \rightarrow F$ & $\sqrt{1/3}(1,-1,-1)$   \\ \hline
 $F \rightarrow D$ & $ \sqrt{1/3}(-1,-1,1)$   \\ \hline
 $D \rightarrow A$ & $\sqrt{1/3}(-1,1,1)$   \\ \hline
$F \rightarrow \overline E$ & $ \sqrt{1/3}(1,-1,-1)$   \\ \hline
 $\overline E \rightarrow \overline A$ & $ \sqrt{1/3}(1,1,-1)$   \\ \hline

\end{tabular}
\vspace{0.5cm}
\end{center}
For both classes of trajectories, the Hamiltonians coincide from $A$ to $F$, from
where the vector $\vec n$ will take a different value for each class of trajectory.

\begin{figure}[h]
\center
\includegraphics[width=4.6in, height=3in]{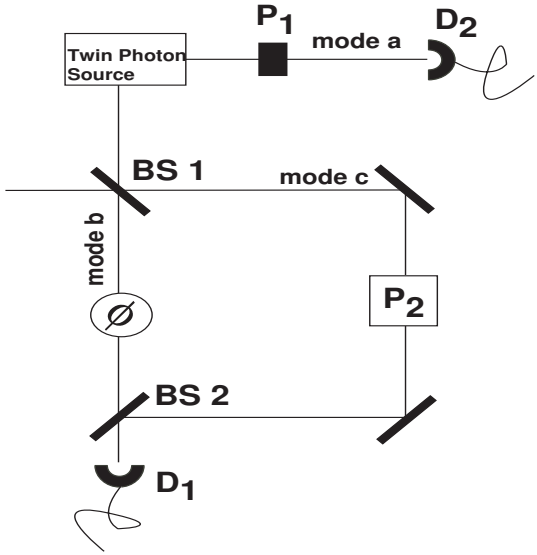}
\caption{Scheme for the experiment enabling the measurement of the global topological phase. Polarization MES are
 produced in the non linear crystal in modes $a$ and $b$. Photons in mode $b$ are combined to mode $c$ in a Mach-Zender
  interferometer. A variable dephasing $\phi$ can be introduced in mode $b$, while wave plates
  properly oriented ($P_1$ and $P_2$) act on arms $a$ and $c$, respectively,  performing the desired trajectories} \label{fig3}
\end{figure}

We will now  address the question of  measuring such an effect in an interference  experiment. In order to test
the topological phase, one needs a setup enabling the interference of an entangled state after performing one of
the trajectories above with a reference one. A simple way to do that consists of interfering entangled
photon pairs,
   as depicted in Fig. (\ref{fig3}). Polarization entangled photon pairs are currently produced with high fidelity and
    efficiency in non linear optical systems \cite{Photons}. Such twin photons propagate in different space modes, which
     will be denoted as $a$ and $b$. Using this notation, the MES state emerging from a non linear crystal is
     $\ket{\Psi}=\sqrt{1/2}\left (\ket{H_a V_b}+\ket{V_a H_b} \right )$. Photons in modes $a$ and $b$ are detected by
     detectors $D_a$ and $D_b$, respectively.  Before detection, photons in mode $b$  enter a Mach-Zender
        interferometer (see Fig.(\ref{fig3})). In the first $50 \%-50\%$ beam splitter (BS 1), mode $b$ is combined to
       mode $c$, which is in the vacuum state , $\ket{0_c}$. After that, the total state is:
\begin{equation}\label{bs1}
\ket{\Psi}=\frac{1}{2}\left ( \ket{H_a V_b 0_c}+i\ket{H_a 0_b V_c}+\ket{V_a H_b 0_c}+i\ket{V_a 0_b H_c}\right ),
\end{equation}
which is exactly what
 is needed  to interfere two entangled states: BS 1 produced two pairs of  polarization MES, each one of them
  entangled to one arm of the interferometer. An experiment  producing the reference fringes consists in
   the introduction of a variable phase factor $e^{i\phi}$ in arm $b$ of the interferometer. Thus,
    before entering the recombining BS (BS 2), the state of the system is, taking into account the dephasing,
    $\ket{\Psi}=\frac{1}{2}\left ( e^{i\phi} \ket{H_a V_b 0_c}+i\ket{H_a 0_b V_c}+e^{i\phi} \ket{V_a H_b 0_c}+i\ket{V_a 0_b H_c}\right )$.
     The passage through BS 2 changes the state into
\begin{eqnarray}\label{bs2}
&&\ket{\Psi}=\frac{1}{2} ( (e^{i\phi}-1)(\ket{H_a V_b 0_c}+i\ket{V_a H_b 0_c})+ \nonumber \\
&&i(e^{i\phi}+1)(\ket{H_a 0_b V_c}+\ket{V_a 0_b H_c}) ).
\end{eqnarray}
Coincidence detections in detector $D_a$ and $D_b$  produce a counting rate $P=1/2|(1-\cos{\phi})|$. The trajectories
one want to implement are realized by photons in modes $a$ and $c$. The space of polarizations, described by the
Poincar\'e sphere, is equivalent to the Bloch sphere \cite{Books}. For polarizations, all rotations in the Poincar\'e
sphere can be accomplished via wave plates properly orientated introducing a dephasing in one of the polarization axis.
The most general rotation matrix can be produced by a sequence of three wave plates:
the first one introducing a dephasing of $\psi$ in the vertical polarization axis, followed by the second one which
introduces a dephasing of $\delta$ and makes an angle $\theta$ with the vertical polarization. Finally, a third plate
introducing a dephasing of $-\psi$ in the vertical polarization can be added. As a result, a rotation matrix is generated
which has exactly the same form as (\ref{evop}) with the following correspondence: the parameter $\delta$ plays the
role of  time $t$, while for the components of vector $\vec n$ we
have $n_x=-\sin{2\theta} \cos{\psi}$, $n_y=\sin{2\theta} \sin{\psi}$ and $n_z=\cos{2\theta}$.
 We can
thus introduce a set of plates acting in mode $a$ and
performing the common part of the trajectories, i.e., from point $A$
to $F$. The remaining of the plates are put in arm $b$ of the
interferometer and their orientation  determines the topology
of the trajectories.  The total state, before reaching BS 2, is then  $\ket{\Psi}=\frac{1}{2}( e^{i\phi}\ket{H_a V_b
0_c}+i(-1)^n\ket{H_a 0_b V_c}+e^{i\phi}\ket{V_a H_b
0_c}+i(-1)^n\ket{V_a 0_b H_c})$. The value of $n$ (odd or
even), depends on the nature of the trajectory. After passing
through BS 2, the whole state transforms into
\begin{eqnarray}\label{bs2b}
&&\ket{\Psi}=\frac{1}{2} ( (e^{i\phi}-(-1)^n)(\ket{H_a V_b 0_c}+\ket{V_a H_b 0_c})+ \nonumber \\
&&i(e^{i\phi}+(-1)^n)(\ket{H_a 0_b V_c}+\ket{V_a 0_b H_c}) ).
\end{eqnarray}
If we now make the same coincidence counts in detectors $D_a$ and $D_b$, we obtain the fringes $P=1/2|((-1)^n-\cos{\phi})|$. If the trajectory is of the {\it minus} class, a dephasing of one half period will be seen in the interference pattern. However, if it is of the {\it plus} type, the fringes remain at the same place. The choice of  splitting the action of the rotations between the first and second qubit was made to stress the interest of using entangled states in the experiment. Since physics shall not depend on the order the detections were made, if we suppose that a detection in $D_a$ is made, the rest of the state can be seen as a statistical mixture of polarizations. This  means that dealing with entangled polarization pairs or with a completely unpolarized beam would produce the same results in our experiment. However, the one-to-one map from photon states to $SO(3)$ elements is only valid for MES and it is only in that case that we can claim probing a topological property of the rotation group.

The gain of the $\pi$ phase factor, as mentioned before, is
related to the number of times the MES crosses the space of its
orthogonal states. An interesting variation of the experiment
could well illustrate a different situation: one could think of
trajectories which are not necessarily cyclic but, having crossed this orthogonal states space, already
correspond to a gain of the $\pi$ phase. This would mean that the
final states interfering in BS 2 will no longer be the same MES.
As a consequence, there will be a decrease in the fringe
visibility, but the desired dephasing would still appear.

Another interesting point to address is the case of states whose entanglement is not maximal. It can
be shown that in that case, the $\pi$ phase also occur. However, the relation between non MES and  the  $SO(3)$
group is more complex \cite{Mosseri:JPA2001}, as well as the geometrical interpretation of
their time evolution.

The authors thank M. Brune, J. M. Raimond and S. Haroche for
patient listening and fruitful comments and  P. H. Souto Ribeiro
for discussion of the experimental aspects. Laboratoire Kastler
Brossel, Universit\'{e} Pierre et Marie Curie and ENS, is
associated with CNRS (UMR 8552).


\end{document}